\documentclass[a4paper]{amsart}
\def\cR{\mathcal{R}}
\def\dd{\mathrm{d}}
\def\del{\partial}
\def\cO{\mathcal{O}}

\usepackage{graphicx}
\usepackage{times}

\title[CSC slices in Schwarzschild]{Constant scalar curvature hypersurfaces in extended
  Schwarzschild space-time}
\date{\today}
\author{M. J. Pareja}
\author{J. Frauendiener}
\address{Institute for Astronomy and Astrophysics, University of
  T\"ubingen, Auf der Morgenstelle 10, 72076 T\"ubingen, Germany}
\email{pareja@tat.physik.uni-tuebingen.de}
\email{joerg.frauendiener@uni-tuebingen.de}

\begin{document}
\begin{abstract}
 We present a class of spherically symmetric hypersurfaces in the
 Kruskal extension of the Schwarzschild space-time. The hypersurfaces
 have constant negative scalar curvature, so they are hyperboloidal in
 the regions of space-time which are asymptotically flat. 
\end{abstract}

\maketitle

\section{Introduction}

Hyperboloidal hypersurfaces have gained some importance in the study
of radiative systems in General Relativity. This is because in a sense
they interpolate between space-like hypersurfaces which become
asymptotically flat and null hypersurfaces which extend out to
null-infinity. Since these hypersurfaces are space-like one can
use them to formulate an initial-value formulation for the Einstein
equations and because they extend to null-infinity one has in
principle full control over the radiation which propagates
outward. One can even go further and compactify the space-time
following the work of Penrose~\cite{penrose63:_asymp_prop} and arrive
at the system of conformal field equations set up by
Friedrich~\cite{friedrich83:_cauch_probl_confor_vacuum_field,friedrich86:_exist_compl}.
For a review on these matters
see~\cite{frauendiener2004:_confor}.

In the asymptotically flat regions of a space-time hyperboloidal
hypersurfaces behave like the space-like hyperboloids in Minkowski
space-time. These are defined by $t^2 - r^2 = \mathrm{const} \equiv
A^{-2}$. They are the only everywhere smooth spherically symmetric
hypersurfaces with constant mean curvature in Minkowski space-time.
They also have constant negative curvature $\cR=-6A^2$. 
When studying the properties of hyperboloidal hypersurfaces in simple
examples like the Schwarzschild and Kruskal space-times we noticed
that much is known for hypersurfaces with constant mean curvature
(see~\cite{brillcavallo1980:_k_schwar,omurchadhamalec2003:_const_schwar}
and references therein), while the class of hypersurfaces with constant
scalar curvature (CSC) seems to have been ignored. Therefore, we decided
to explore this class of hypersurfaces in some detail.

We find that hypersurfaces in this class are characterised by three
constants: the scalar curvature, the location of the hypersurface in
space-time (i.e.~an event in space-time lying on the hypersurface), 
and an integration constant. The hypersurfaces can be
space-like or time-like. The latter ones either connect the white-hole
singularity with the black-hole singularity or they 'stop' with a
diverging extrinsic curvature somewhere inside the white-hole. Most
space-like hypersurfaces connect one asymptotic region of the Kruskal
extension with the other one. But they all remain below the
bifurcation sphere. There is another class of space-like hypersurfaces
which come in from null-infinity and reach a minimal radius where the
extrinsic curvature diverges.

The paper is organised as follows. In sect.~\ref{sec:initialdata} we
solve the constraints for a hypersurface with constant negative scalar
curvature. In sect.~\ref{sec:embedding} we discuss the embedding
equation for such hypersurfaces in Schwarzschild and derive the
qualitative behaviour of the embedded slices. In
sect.~\ref{sec:kruskal} we look at the behaviour of the hypersurfaces
in the Kruskal extension of the Schwarzschild space-time and close the
paper with a short discussion of our findings.

\section{Initial data sets with constant scalar curvature}
\label{sec:initialdata}

In this first section we want to determine spherically symmetric
initial data $(g_{ab},K_{ab})$ on a 3-dimensional
hypersurface~$\Sigma$ with constant scalar curvature. In our
discussion we follow closely the treatment
of~\cite{omurchadhamalec2003:_const_schwar}. First we determine the
spherically symmetric metrics with constant scalar curvature and then
proceed to solve the vacuum constraints for the extrinsic curvature.

\subsection{3-metrics with constant scalar curvature}
\label{sec:metrics}

We are interested in spherically symmetric 3-dimensional metrics with
constant negative scalar curvature $\cR$ on a 3-manifold $\Sigma$.
These can be obtained directly by solving the equation
\[
  \cR = -6A^2 = const
\, ,
\]
where $A$ is an arbitrary positive constant on $\Sigma$.
The general spherically symmetric metric on a 3-dimensional manifold
can be written in the form
\[
g = \dd s^2 + f(s)^2\,\dd\sigma^2
\, ,
\]
where $s$ is the geodesic radial distance on the 3-manifold and
$\dd\sigma^2$ is the metric on the unit 2-sphere. The function $f$
determines the radius $r$ of the spheres of symmetry and hence determines
the intrinsic geometry. We find
the scalar curvature for this metric and obtain the equation
\[
\cR = -6A^2 = -4 \frac{f''}{f} - \frac{2f'^2}{f^2} + \frac{2}{f^2}
\, .
\]
Multiplying with $f'$ and rearranging we obtain the equation\footnote{We ignore
  here the case $f'=0$.}
\[
(1+3A^2 f^2)f' = 2ff'f'' + f'^3 = (ff'^2)'
\, ,
\]
which can be integrated once 
\[
f + A^2f^3 = ff'^2 + 2C
\]
with an integration constant $C$. Then we can write
\[
f'^2 = 1 - \frac{2C}{f} + A^2 f^2 
\]
and this equation cannot be integrated in closed form any
further. The explicit solution involves elliptic integrals and is not
very illuminating. However, writing $r=f(s)$ and introducing $r$ as a new
coordinate, the metric takes the form
\begin{equation}
g = \frac{\dd r^2}{1 - \frac{2C}{r} + A^2 r^2} + r^2 \dd
\sigma^2
\, .
\label{eq:metricinr}
\end{equation}
Clearly, rescaling the metric with a constant $\lambda$ so that $g
\mapsto \lambda^2 g$ rescales the scalar curvature $\cR \mapsto
\cR/\lambda^2$ so that $A\mapsto A/\lambda$. This can also be seen
from~\eqref{eq:metricinr} which, in addition, reveals that $C\mapsto
\lambda C$. Note that the transition to this
form~\eqref{eq:metricinr} of the metric is valid only as long as
$f'(s)\ne0$.

\subsection{Solving the constraints}
\label{sec:constraints}

We consider now the vacuum constraints.  They are written as
\begin{gather*}
  \cR - K_a{}^bK_b{}^a + (\mathrm{tr}K)^2 = 0 \, ,\\
  \del_a K_b{}^a - \del_b \mathrm{tr}K = 0
\, .
\end{gather*}
Here $\del_a$ denotes the covariant derivative compatible with the
metric $g_{ab}$ and $\mathrm{trK}=K_c{}^c$.
Using the spherical symmetry we find that the extrinsic curvature
necessarily has the form
\[
K_{ab} \, \dd x^a\dd x^b = \tilde k(r)\,\dd r^2 + \tilde \kappa(r)\, \dd\sigma^2
\, ,
\]
for two functions $\tilde k(r)$ and $\tilde \kappa(r)$. Therefore, the
$(1,1)$-tensor $K_a{}^b$ is diagonal in the given coordinate basis
with components
\begin{equation}
  \label{eq:Kab}
  K_a{}^b \doteq \mathrm{diag}(k(r),\kappa(r),\kappa(r))
\, .
\end{equation}
Hence, the trace of the extrinsic curvature is
\[
\mathrm{tr}K = k+2\kappa =: 3K
\, ,
\]
thus defining the mean curvature $K(r)$. 
Expressing $\kappa$ in terms of $K$ and $k$ as
\[
\kappa(r) = \frac12 \left(3K(r)-k(r)\right)
\]
and inserting into the constraints we have
\begin{gather}
  \cR - \left( k^2 + \frac12 \left( 3K - k \right)^2 \right) +
  9K^2 = 0 \, , \label{eq:hamilton}\\
  r \del_r (k-3K) = -3(k-K)
\, .
\label{eq:momentum}
\end{gather}
Here the metric has a constant negative scalar curvature $\cR$,
which we write as $\cR=-6A^2$ as before. Rescaling the metric with a
constant $\lambda^2$ results by virtue of the constraint equations in
a rescaling of the extrinsic curvature $K_{ab}$ with $\lambda$. In
other words, when $(g_{ab},K_{ab})$ is a solution of the
constraints, then $(\lambda^2 g_{ab}, \lambda K_{ab})$ is also a
solution.

We first solve the Hamiltonian constraint for the function~$k$
\begin{equation}
k = K \pm 2 \sqrt{K^2 - A^2}\label{eq:k}
\end{equation}
and insert this into the momentum constraint. We get the equation
\begin{equation}
A^2 rK' = -3 (K^2-A^2) \left( K \pm \sqrt{K^2 - A^2}\right)
\, .
\label{eq:Kprime}
\end{equation}
This equation has the solution
\begin{equation}
  \label{eq:K_r}
  K(r) = \mp A \frac{r^3 + D^3}{\sqrt{r^3(r^3+2 D^3)}}
\, ,
\end{equation}
for some constant of integration $D$. For $D>0$ the mean curvature is
defined for all values of $r$, while for $D<0$ there is only a limited
range given by $r>-2^{1/3}D$. This means that the hypersurface has to
stop somewhere because the extrinsic curvature blows up. For $D=0$ we
have constant mean curvature.  From~\eqref{eq:k} we find
\begin{equation}
  \label{eq:k_r}
  k(r) = \mp A \frac{r^3 - D^3}{\sqrt{r^3(r^3+2D^3)}}
\end{equation}
and, finally,
\begin{equation}
  \label{eq:m_r}
  \kappa(r) = \mp A \sqrt{1 + 2 (D/r)^3}
\, .
\end{equation}
The ambiguity in sign is inherent in the vacuum constraints because with
$K_{ab}$ also $-K_{ab}$ is a solution of the constraints. It
corresponds to the choice of the time orientation. We will concentrate
in the following on the $+$ sign, the case where the  mean
curvature is positive. Then $K(r)>A$ and approaches $A$ for $r\to\infty$.

So far we have not considered the fact that the hypersurface should
be embedded in the Schwarzschild space-time. This can be taken into
account by the consideration of the Misner-Sharp
mass~\cite{misnersharp1964:_relat}\footnote{We are grateful to N. O'Murchadha for
pointing this out to us.}. This mass $m(r,t)$ is defined in a
spherically symmetric space-time by the equation
\[
g_{rr} = \left(1-\frac{2m(r,t)}{r}+U(r,t)^2 \right)^{-1},
\]
where, in our case, $U(r,t)=r \kappa(r)$, and it has the property that
in the Schwarzschild space-time it is constant, equal to the mass
parameter $m$. Inserting $g_{rr}$ from~\eqref{eq:metricinr} and
$\kappa(r)$ from above we obtain a relation between the integration
constants $D$, $C$, and $A$
\begin{equation}
m-C = A^2 D^3.\label{eq:constants}
\end{equation}
In particular, substituting into \eqref{eq:K_r},
the mean curvature in terms of $m$ and $C$ is
\begin{equation} 
  \label{eq:K_rC}
  K(r) = \frac{A^2 r^3 + m - C}{\sqrt{r^3(A^2 r^3 + 2(m - C))}}
\, .
\end{equation}
Remarkably, the case $C = m$, in which $K(r) = A$ and also $k(r) =
\kappa(r) = A$, characterises a hypersurface with constant mean
curvature and constant scalar curvature, for which the extrinsic
curvature is `pure trace' $K_{ab} = A \, g_{ab}$.

\section{Embedding CSC slices in Schwarzschild space-time}
\label{sec:embedding}

In the next two sections we ask how hypersurfaces with the intrinsic
and extrinsic geometries determined in the previous sections can be
embedded into the Schwarzschild space-time. We first discuss this
question in the well-known chart of Schwarzschild coordinates and
discuss the qualitative behaviour of the hypersurfaces. Then we write
down a system of equations for use in the Kruskal completion of the
Schwarzschild space-time.

\subsection{Schwarzschild coordinates}
\label{subsec:schwarzschild}

We consider the Schwarzschild space-time and write its metric in the
usual Schwarzschild coordinates $t$ and $r$. However, in order to
simplify the formulae we rescale the coordinates with $2m$, i.e. we
measure them in units of the Schwarzschild radius. Then the areal
radius is $2mr$. In these coordinates the metric is
\[
\dd s^2 = 4m^2 \left\{- \left(1-\frac1r\right) \dd t^2 +
\left(1-\frac1r\right)^{-1} \dd r^2 + r^2 \dd\sigma^2 \right\}
\, ,
\]
which is valid for $0<r<1$ or $r>1$. We locate a spherically symmetric
hypersurface by using a height-function $h$ with an equation of the
form
\[
t = h(r)
\, .
\]
For such a hypersurface one can find the intrinsic metric, the normal
vector, and the extrinsic curvature expressed in terms of
derivatives of the height function. The intrinsic metric is
\begin{equation}
  \label{eq:intrinsic_metric}
  g = 4m^2 \left\{F\, \dd r^2 + r^2 \dd\sigma^2 \right\}
\, ,
\end{equation}
with
\[
F = \frac{r}{r-1} - \frac{r-1}{r}\,h'(r)^2
\, .
\]
Notice, the hypersurface will be space-like (resp. time-like, null)
whenever $F>0$ ($F<0$, $F=0$). Defining $N=1/\sqrt{|F|}$,
the normal unit vector is 
\begin{equation}
  \label{eq:normal_vector}
  n =  \frac{N}{2m}\left(\frac{r}{r-1}\, \del_t + h'
  \frac{r-1}{r}\,\del_r\right)
\, .
\end{equation}
This vector is chosen such that it is future-pointing whenever $F>0$. From this
expression we compute the extrinsic curvature
\begin{equation}
  \label{eq:extrinsic_curvature}
  K_{ab} \, \dd x^a \dd x^b = \frac{2m}{N^2} \left(
    Nh'\left(1-\frac1r\right)\right)'\,\dd r^2 
  + 2m r \left(Nh'\left(1-\frac1r\right)\right)\, \dd\sigma^2
\end{equation}
and obtain for the function $k$ and $K$ the following expressions
\[
k(r) = \frac1{2m}\, \del_r\left(Nh' \left( 1 - \frac1r \right)\right),\qquad 
K(r) = \frac1{6mr^2}\, \del_r \left( r^2 Nh'\left(1-\frac1r \right)\right)
\, .
\]
We have two conditions on the height-function $h$: $(i)$ the intrinsic
metric should be of the form~\eqref{eq:metricinr} and $(ii)$ the mean
curvature should be given by~\eqref{eq:K_r} with the positive
sign. From condition $(i)$ we find, after adjusting the $r$ coordinate
and redefining the constants $a:=2mA$ and $c:=C/m$, that
\begin{equation}
1-\frac{c}{r} + a^2r^2 = \frac1F\label{eq:metric_equality}
\end{equation}
holds. Solving this equation for $h'$ yields
\[
h'^2 = \frac{r^2}{(r-1)^2} \frac{1 + a^2 r^3 - c}{r+a^2r^3 - c}
\, .
\]
Condition $(ii)$ implies with $d=D/2m$
\[
\frac1{3r^2} \del_r
\left( r^2 Nh'\left(1-\frac1r \right)\right) = 
a\frac{r^3 + d^3}{\sqrt{r^3(r^3+2d^3)}}
\, ,
\]
which can be integrated once to the equation
\[
r Nh'(r-1) = a\sqrt{r^3(r^3+2d^3)}
\, .
\]
This can be transformed into an expression for $h'$
\[
h'^2 = \frac{r^2}{(r-1)^2} \frac{a^2 r^3 + 2 a^2 d^3}{r + a^2r^3
  +2a^2d^3 - 1}
\, .
\]
Thus, the two conditions on $h'$ agree if and only if 
\[
2a^2d^3 = 1-c \, , \ \ \text{ i.e. } \, \ A^2 D^3 = m - C
\, ,
\]
i.e. if and only if the relation~\eqref{eq:constants} between the
integration constants holds.

For large $r$ we get $h'^2 \to 1$ and from the two branches of $h'$ we
restrict ourselves to the positive sign for which $h'\to1$. In
this case, the hypersurface extends out to future null-infinity. So we
define
\[
Q(r) := \sqrt{\frac{1 + a^2 r^3 - c}{r+a^2r^3 - c}}
\, .
\]
Then we have the embedding equation
\begin{equation}
  \label{eq:hprime}
  h'(r) = \frac{r}{r-1} Q(r).
\end{equation}

\subsection{The surfaces of constant radius}
\label{subsec:constrad}

Clearly, in the Schwarzschild space-time, where the only relevant
coordinate is the radial coordinate $r$, the hypersurfaces given by
$r=const \,$ are necessarily also hypersurfaces of constant scalar
curvature. These hypersurfaces cannot be obtained by using a
height-function. They are time-like outside the horizon and space-like
inside.

\subsection{Regularity at the horizon}
\label{subsec:horizon}

Obviously, \eqref{eq:hprime} is singular at the horizon $r=1$. But,
since the singularity at $r=1$ is only a coordinate singularity, this
does not imply that the hypersurface itself ends at the horizon. This
can be seen in several ways. We switch to a different time coordinate
$\bar t = t - \log(r-1)$ so that the Schwarzschild metric takes
the form
\[
\dd s^2 = 4m^2 \left\{-\left(1-\frac1{r}\right) \dd \bar t^2 -
\frac{2}{r}\,\dd \bar t \dd r + \left(1+\frac1r\right)\, \dd r^2
+ r^2 \dd\sigma^2 \right\},
\]
which is manifestly regular at the horizon. With this time coordinate
the hypersurface is given by the equation
\[
\bar t = \bar h(r) = h(r) - \log(r-1)
\]
and we get
\[
\bar h'(r) = h'(r) - \frac1{r-1} = \frac{r Q(r) - 1}{r - 1}
\, .
\]
Except for the special case when $a^2=c-1$, the function $Q(r)$ is
smooth at $r=1$ with $Q(1)=1$. Hence, the function $\bar
h(r)$ is smooth as well at $r=1$, so that the hypersurfaces extend
smoothly across the horizon. The exceptional case $a^2=c-1$ will be
discussed below.

\subsection{The embedding equation}
\label{subsec:poss}

Knowing that the hypersurfaces are smooth across the horizon we
restrict ourselves in our discussion to the usual Schwarzschild
coordinates. The crucial object in our consideration are the
polynomials
\[
p(r) := r + a^2r^3 - c \quad \text{and}\quad q(r) := p(r)-r+1
\, ,
\]
in terms of which we get
\[
Q^2(r) = \frac{q(r)}{p(r)}
\, .
\]
The behaviour of the hypersurfaces is characterised by the zeros of
numerator and denominator of $Q^2(r)$. Since both polynomials are
increasing with $r$ they can have at most one zero. From the
metric~\eqref{eq:metricinr} it follows that the dependence of the
areal radius on the geodesic distance is given by the equation
\[
\frac{\dd r}{\dd s} = \sqrt{\frac{|p(r)|}{r}}
\, .
\]
Therefore, the zero of $p$ is the extremal value of the area of the
spheres of symmetry. The sign of $p(r)$ decides about the causal
character of the hypersurfaces. For positive values the hypersurfaces
are space-like and for negative values they are time-like.

Since at these values the second derivative of $r$ with respect to $s$
is
\[
\frac{\dd^2 r}{\dd s^2} = \mathrm{sign}(p(r))\frac{1+3a^2r^2}{2r}
\, ,
\]
these are minimal surfaces when the hypersurfaces are space-like and
they are maximal surfaces when they are time-like. We notice, at the
extremal point the hypersurface is tangent to the hypersurface of
constant $r$ through that point. It follows from
equation~\eqref{eq:Kprime} that $K'(r)$ is negative for all values of
$r$. So $K(r)$ is decreasing with $r$, which implies that $K$ is
maximal at the extremal point in the space-like case, while it is
minimal in the time-like case.

At the zero of the numerator of $Q^2(r)$ we have 
\[
2m (1 - c + a^2 r^3) = A^2(2D^3+r^3) = 0
\, .
\]
Comparison with~\eqref{eq:K_r},~\eqref{eq:k_r}, and \eqref{eq:m_r}
shows that these are the locations where the extrinsic curvature blows
up. This implies that the surfaces stop at this radius because they
can no longer be embedded into the Schwarzschild space-time.

The extremal surface is located at the point with $p(r)=0$, while
the extrinsic curvature diverges at the point with $p(r) = r-1$. The
location of these points is determined by the values of the constants
$a$ and $c$. The graph of the polynomial $p(r)$ always intersects the
abscissa with slope $p'(0)=1$. The value at the intersection is
$p(0)=-c$.  Since $p$ grows unboundedly for large $r$ with a rate
given by $a^2$ and since $c$ determines the intersection at $r=0$, we
have generically three main cases (the non-generic special cases will
be discussed in the next section) characterised by the intersection
points of the polynomial with the line $L=\{(r,r-1):r\ge0\}$. Of
course, the hypersurfaces can exist only in regions in which $p(r)$ and
$q(r)$ have the same sign, i.e. when the polynomial takes positive
values and lies above $L$ or when the polynomial takes negative values
and lies below $L$. The various cases are
\begin{figure}[htb]
  \centering
  \includegraphics[width=8cm]{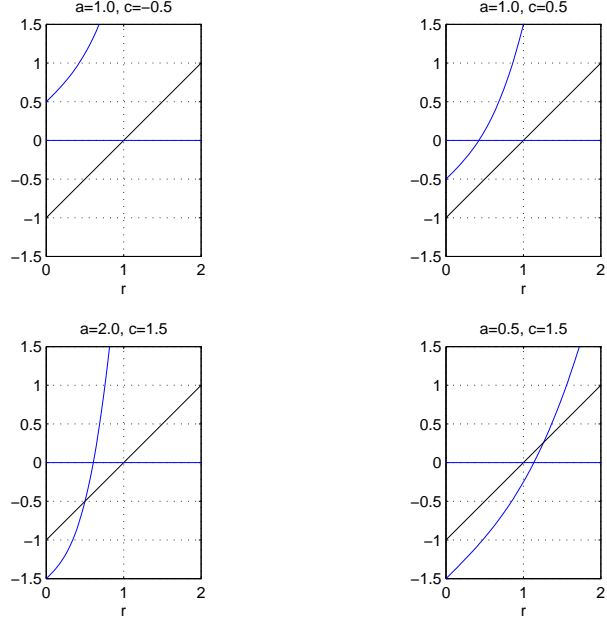}
  \caption{The polynomial $p(r)$ and its relation to the line $L$
    for different values of $a$ and $c$.}
  \label{fig:poly}
\end{figure}
\begin{enumerate}
\item $c<0$. In this case $p(r) > r-1$ and $p(r) > 0$ for all $r\ge0$,
  so there are no extremal surfaces and no singularities. The surfaces
  exist for all $r$ and they are space-like.
\item $0<c<1$. Again, $p(r)>r-1$ but $p(r)$ becomes negative for small
  values of $r$. There is a zero at a value $r_0 \le c < 1$. So the
  surfaces in this case have a minimal surface but no singularity. The
  hypersurfaces are space-like. They come in from null-infinity reach
  a minimal hypersurface at $r_0$ and then extend out again to
  null-infinity. 
\item $c>1$. This is a more complicated case. For all values of $a$
  the polynomial $p$ has a zero at a value $r=r_0$ and intersects the
  line $L$ at a value $r=r_1$. We have to distinguish two subcases:
  \begin{enumerate}
  \item $1<c<1+a^2$. In this case $r_1 < r_0 < 1$. The hypersurfaces exist in
    the region $0 \le r \le r_1$ where they are time-like and $r_0 \le r$
    where they are space-like. The
    time-like hypersurfaces reach the Schwarzschild singularity
    at $r=0$ and grow towards larger radii but stop at $r_1< 1$ where
    their extrinsic curvature diverges. The space-like hypersurfaces 
    come in from null-infinity and,
    crossing the horizon, reach a minimal surface with radius $r_0$.
    Then they go out again to null-infinity. 
  \item $1+a^2<c$. Here $1<r_0 < r_1$ and the two regions for existence
    of the hypersurfaces are reversed. The hypersurfaces on $0 \le r \le r_0$
    extending from the singularity are time-like, grow towards a surface of
    maximal radius and then shrink back to $r=0$. The outer
    hypersurfaces on $r_1 \le r$ are space-like, come in from null-infinity
    and end at $r_1$ with a diverging extrinsic curvature.
  \end{enumerate}
\end{enumerate}

\section{Embedding CSC slices in Kruskal space-time}
\label{sec:kruskal}

The Kruskal extension of the Schwarzschild space-time is covered by
coordinates $X$ and $T$ which are related to the Schwarzschild
coordinates by the relations
\begin{equation}
  \label{eq:kruskal_coord}
  e^r(r-1) = X^2 - T^2,\qquad
  t = \log\left(\frac{X+T}{X-T}\right).
\end{equation}
In these coordinates the metric takes the form
\[
\frac{4 e^{-r}}{r} \left(-\dd T^2 + \dd X^2\right) + r^2 \dd\sigma^2,
\]
where $r$ has to be regarded as a function of $X^2-T^2$
via~\eqref{eq:kruskal_coord}.

\subsection{The embedding equation}
\label{subsec:embeq}

To obtain the embedding equation in terms of the Kruskal coordinates
we write $r$ and $t$ as functions of some parameter $s$, so that $t(s)
= h(r(s))$, and then regard also $T$ and $X$ as functions of $s$. Using
the relationship between the coordinates and the embedding equation
one derives the equations
\begin{align*}
  \dot T &= \frac12 \left(X h'(r) + T r/(r-1)\right) \dot r, \\
  \dot X &= \frac12 \left(T h'(r) + X r/(r-1)\right) \dot r.
\end{align*}
With $h'(r) = Q(r) r/(r-1)$ one derives easily the equation
for the height function $T=T(X)$ as
\begin{equation}
\frac{dT}{dX} = \frac{Q X + T}{X + Q T},
\label{eq:dTdX}
\end{equation}
from which it is obvious, that the horizon at $r=1$ is not a
concern. From this form of the embedding equation one quickly derives
the following symmetry properties of the solutions. If $T(X)$ is a
solution of the equation with $Q=\sqrt{q/p}$ taken from the positive
branch of the square root, then $-T(X)$ is a solution for
$-Q$. Similarly, if $T(X)$ is a solution for positive $Q$, then 
$\bar T(X)=T(-X)$ is a solution for $-Q$.

In order to derive further properties we write down the embedding
equation in parametric form by choosing $s/2$ as the geodesic distance
along the radial directions and we arrive
at the system
\[
\begin{aligned}
  (r-1)\, \dot T &= \left( \sqrt{r|q(r)|} X + \sqrt{r|p(r)|} T\right) ,\\
  (r-1)\, \dot X &= \left( \sqrt{r|q(r)|} T + \sqrt{r|p(r)|} X\right) .
\end{aligned}
\]
In terms of the null coordinates $U=T-X$ and $V=T+X$ this system can
be written as
\begin{equation}
\begin{aligned}
  (r-1)\, \dot U &= \left( \sqrt{r|p(r)|} - \sqrt{r|q(r)|} \right) U,\\
  (r-1)\, \dot V &= \left( \sqrt{r|p(r)|} + \sqrt{r|q(r)|} \right) V .
\end{aligned}\label{eq:UVsystem}
\end{equation}
We list some of the obvious properties of the system:
\begin{enumerate}
\item In the asymptotic region $r\to\infty$ we have 
  \[
  \dot U \approx 0,\qquad \dot V \approx 2ar^2\,V,
  \]
  which implies that the hypersurfaces approach the light cones
  $U=const$ for large~$r$.
\item Similarly, for large values of the parameter~$a$ the
  hypersurfaces approach null hypersurfaces. This can easily be seen
  already from the form of the metric~\eqref{eq:metricinr}. With
  $g_{rr}=\frac{r}{p(r)}$ it is clear that the metric degenerates for
  $a\to\infty$.
\item Note that the case $h'(r)\to1$ for $r\to\infty$ referred to in
  sect~\ref{subsec:schwarzschild} corresponds to a hypersurface which
approaches future null-infinity for large values of $X$, while the
case $h'(r)\to-1$ gives a hypersurface approaching future
null-infinity in the other asymptotic region of the Kruskal extension
where $X<0$. This implies that for hypersurfaces connecting both
null-infinities we have to consider $Q$ and $-Q$.
\item Changing the branch of $\sqrt{r|q(r)|}$ corresponds to the
  interchange of $U$ and~$V$, i.e. to $X\mapsto -X$.  Similarly, the
  system is invariant under the transformation $(U,V)\mapsto
  (-U,-V)$. These transformations together correspond to the $T\mapsto
  -T$.  So we can restrict ourselves to consider only hypersurfaces which
  lie in the region $U<0$, corresponding to the regions~I and~III in the
  standard Kruskal extension.\footnote{We use the labeling of
    Wald\cite{wald84:_gener_relat} for the four
  regions of the Kruskal extension. }
\item At the extremal surfaces given by $p(r)=0$ we have 
\[
\frac{\dot U}{U} = -\frac{\dot V}{V} \, , \qquad
\frac{\dot T}{\dot X} = \frac{X}{T}
\, .
\]
This means that at these points $\frac{d(UV)}{ds}=0$, so that the hypersurfaces
touch the hypersurface $r=const \,$ at these points.
\item At the locations with $q(r)=0$, where the extrinsic curvature of
  the hypersurfaces diverges, we have
\[
\frac{\dot U}{U} = \frac{\dot V}{V} \, , \qquad
\frac{\dot T}{\dot X} = \frac{T}{X}
\, .
\]
In this case, the condition implies that $\frac{d(T/X)}{ds} =0$, i.e. the
hypersurfaces touch the hypersurface $t=const \,$ through that point. 
\end{enumerate}

\subsection{Behaviour at the horizon}
\label{subsec:horizonkruskal}

The horizon at $r=1$ corresponds to the two null lines in the Kruskal
extension defined by either $U=0$ or $V=0$. In order to see
the behaviour of the system close to these lines we need to expand the
function $r(U,V)$ for small $U$, respectively $V$. 
We make the discussion for the case where $p$ and $q$ are positive, 
an analogous argument applies to the case where they are negative.
We have
\[
y = -U V = e^r(r-1) = e \left( (r-1) + (r-1)^2 + \frac12 (r-1)^3 + \cO(r-1)^4
  \right)
\, ,
\]
which can be inverted to
\[
(r-1) = (y/e) - (y/e)^2 + \frac32 (y/e)^3 + \cO(y^4)
\, .
\]
Furthermore, near $r=1$ we have (in the generic case $\sqrt{1 + a^2 -
  c}\ne0$)
\[
\begin{gathered}
\sqrt{rq(r)}  - \sqrt{rp(r)} = \frac{r-1}{2\,{\sqrt{1 + a^2 - c}}} + 
  \frac{1 - 4\,a^2 - 2\,c}{8\,
     {\left( 1 + a^2 - c \right) }^{\frac{3}{2}}} \;(r-1)^2 + 
  {\cO(r-1)^3},\\
  \sqrt{rq(r)}  + \sqrt{rp(r)} = 2\,{\sqrt{1 + a^2 - c}} + 
  \frac{3 + 8\,a^2 - 2\,c }{2\,{\sqrt{1 + a^2 - c}}}\; (r-1) + 
  {\cO(r-1)}^2.
\end{gathered}
\]
Thus, near $r=1$ the system can be written as
\[
\begin{aligned}
  \dot U &= \frac{U}{2\,{\sqrt{1 + a^2 - c}}} - 
  \frac{1 - 4\,a^2 - 2\,c}{8e\,
     {\left( 1 + a^2 - c \right) }^{\frac{3}{2}}} \;U^2V + \cO(U^2V^2) \, , \\
  \dot V &= -\frac{2e\,{\sqrt{1 + a^2 - c}}}{U} + 
  \frac{7 + 12\,a^2 - 6\,c }{2\,{\sqrt{1 + a^2 - c}}}\; V + 
  {\cO(UV)} 
\, .
\end{aligned}
\]
Clearly, in the neighbourhood of $V=0$ the system is regular and has
regular solutions. The hypersurfaces determined by the system cross
the horizon $V=0$ in a regular way. However, the horizon at $U=0$ is
different. In this case there is no way to approach the horizon
because the equation becomes singular at $U=0$.

The behaviour at the two horizons $U=0$, i.e. $T=X$, and $V=0$, i.e.
$T=-X$, can be obtained directly from equation~\eqref{eq:dTdX}. Since
$Q$ is positive, we find that $T=X$ is a solution, while $T=-X$
is not a solution. This implies, by uniqueness of solution, that $T=X$
cannot be crossed by other solutions, while $T=-X$ can be crossed.

\subsection{Discussion of the different cases}
\label{subsec:cases}

Let us now discuss the different cases found in the previous section
and determine the generic behaviour of the hypersurfaces. 
\subsubsection{The case $c<0$}
In this case $p$ and $q$ have no zeros and $\sqrt{r p(r)} > \sqrt{r
  q(r)}$ for $r>1$, while $\sqrt{r p(r)} < \sqrt{r q(r)}$ for $r<1$.
Thus, it is easy to see that $\dot U >0$ for $U<0$ and $\dot V>0$ for
$V>0$. This implies that a hypersurface which happens to get into
region~I where $U<0$ and $V>0$ will approach $U=0$ for large positive
values of $s$, while in the other direction it will approach $V=0$ and,
because the equation is regular there, it will cross into region~III,
the white hole region, where $V<0$ and hence $\dot V>0$. So, in this
case, the hypersurfaces emerge from the white hole and extend all the
way out towards null-infinity.

\begin{figure}[htb]
  \centering
  \includegraphics[width=8cm]{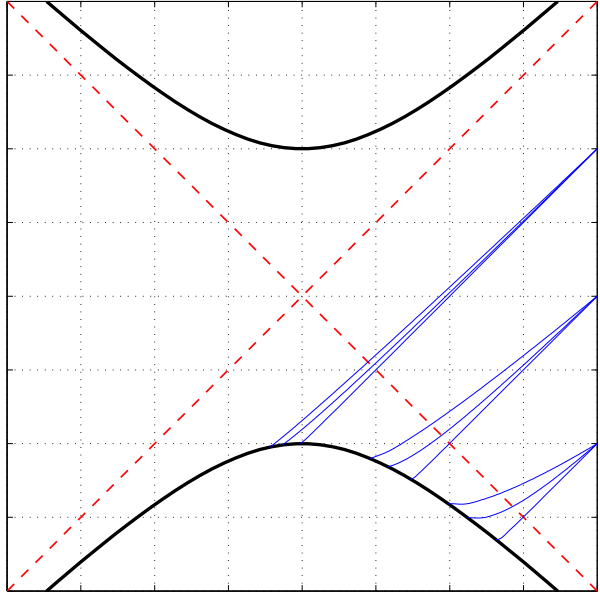}
  \caption{
  \label{fig:case1}
  Three classes of hypersurface in different regions of space-time for
the case $c=-0.5$. Within each class the parameter $a$ takes the
values $a=10$, $a=1$, and $a=0.01$.}
\end{figure}

In Fig.~\ref{fig:case1} we show hypersurfaces for $c=-0.5$. We exhibit
three different classes of hypersurfaces. Within each class there are
three hypersurfaces with different values of $a$ which go through the
same space-time event. The values of $a$ are $a=10$, $a=1$, and
$a=0.01$. It can be clearly seen that the large values of $a$ result
in an almost null hypersurface.

\subsubsection{The case $c=0$}

This case is almost the same as the case $c<0$. Also in this case the
hypersurfaces from any space-time event emerge from the singularity at
$r=0$ except that now the location $r=0$ is also a point where $\dot
r = 0$, so it is a minimal surface. The hypersurfaces run into the
singularity tangentially.

\subsubsection{The case $0<c<1$}

In this case the polynomial $p(r)$ vanishes at $r=r_0$ between $r=0$
and $r=1$.  So a corresponding hypersurface has a minimal surface.  At
that minimal surface the geometry of the hypersurface is completely
regular. The hypersurface which is space-like comes in from
null-infinity at large positive values of $X$, and touches the
hypersurface of constant radius $r=r_0$ in a minimal surface. TYhere
the solution switches the branch and continues on into region~IV where
it ultimately reaches the other future null-infinity. Along the way it
smoothly crosses both horizons. Since one branch of the solution only
covers the part of the hypersurface up to the minimal surface, the
complete hypersurface is obtained by switching the branch of
$\sqrt{p(r)}$ at the minimal surface. All the hypersurfaces in this
case connect one null-infinity with the other.  They all lie below the
throat at $\, T=0=X$.

\begin{figure}[htb]
  \centering
  \includegraphics[width=8cm]{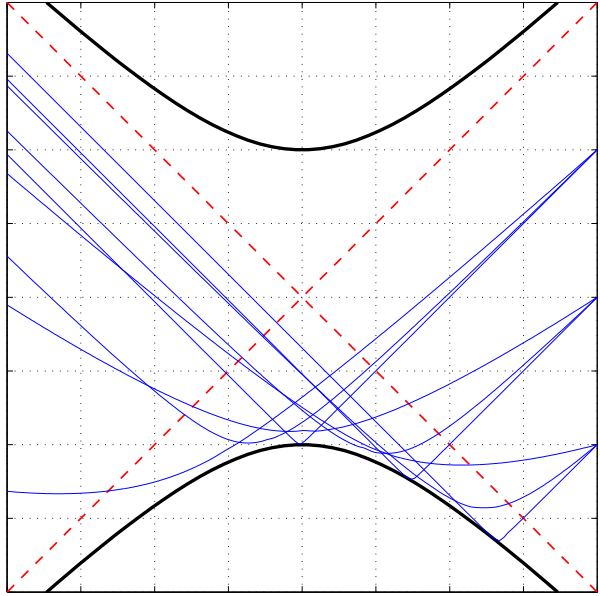}
  \caption{The case $0<c<1$}
  \label{fig:case2}
\end{figure}

Again, for large values of $a$ the hypersurfaces approach
null-hypersurfaces.  In this case these are two null-hypersurfaces
intersecting in the minimal sphere at $r=r_0$ outside the singularity.
In Fig.~\ref{fig:case2} we show again a sample of hypersurfaces with
$c=0.5$ and the same values for $a$ and initial conditions as in
Fig.~\ref{fig:case1}.

\subsubsection{The case $c=1$}

In this case, as in the previous case, the polynomial $p(r)$ vanishes
between $r=0$ and $r=1$. The additional feature is the emergence of a
point where $q(r)$ vanishes. For the special case $c=1$ treated here
this location is at the singularity. The hypersurfaces with $a\ne0$
have the same qualitative behaviour is the ones treated in the
previous case. This case yields hypersurfaces with constant mean
curvature and constant scalar curvature (considered by
Schmidt~\cite{schmidt2002:_data_krusk}, see also Brill et
al.~\cite{brillcavallo1980:_k_schwar}) for which the extrinsic
curvature is pure trace $K_{ab}= K g_{ab}$.

The case $a=0$ is interesting because then $q$ vanishes identically so
that the solutions of the system are
\[
T/X = const
\, ,
\]
which are the asymptotically euclidean hypersurfaces $t=const \,$ of
the Schwarzschild space-time.

\subsubsection{The case $c>1$}

\begin{figure}[htb]
  \centering
  \includegraphics[width=8cm]{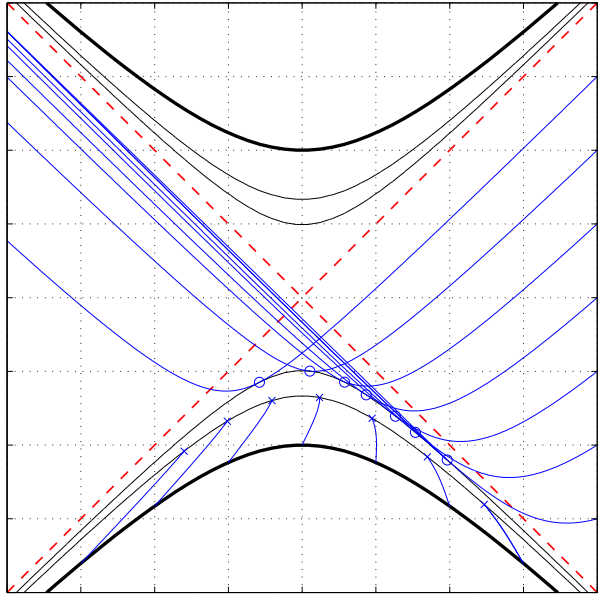}
  \caption{The case $1<c<1+a^2$. We show the hypersurfaces for fixed
    values of $a$ and $c$, varying their location in space-time.}
  \label{fig:case3a}
\end{figure}

Now the polynomial $p(r)$ starts out below the line $L$ and (for all
non-zero values of $a$) there is an intersection with this line at
$r=r_1$. This intersection may lie below $r=1$ or above $r=1$
depending on the parameters $a$ and $c$. The critical case is given by
$a^2=c-1$. We first discuss the case $1<c<1+a^2$. Then the
intersection is inside $r=1$ and the polynomial hits the $r$-axis at a
value $r=r_0 > r_1$ after intersecting the line. Hypersurfaces exist
for all values of $r\ge r_0$. Again, the geometry at the minimal
surface $r=r_0$ is regular so we can continue the hypersurface by
switching to the other branch of $\sqrt{p(r)}$. We get again the same
qualitative behaviour of the hypersurfaces. They connect one
null-infinity with the other. Some of the hypersurfaces are shown in
Fig.~\ref{fig:case3a}.
The other region of existence for hypersurfaces is inside the horizon.
These hypersurfaces are time-like. They emerge from the singularity
and grow towards the future until they come to an end when $q$
vanishes, i.e. when the extrinsic curvature diverges.

\begin{figure}[htb]
  \centering
  \includegraphics[width=8cm]{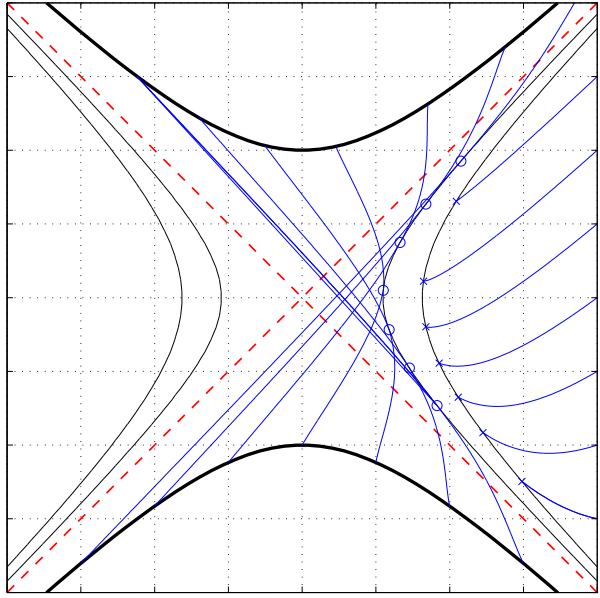}
  \caption{The case $1+a^2<c$. Hypersurfaces for fixed values of $a$
    and $c$, and varying location in space-time.}
  \label{fig:case3b}
\end{figure}

The other case is when $1+a^2<c$. In this case we have space-like
hypersurfaces in region~I which end with a diverging extrinsic
curvature, see Fig.~\ref{fig:case3b}, and there are time-like
hypersurfaces which come out of the white hole singularity, grow up to
a maximal surface and then, after switching to the other branch of
$\sqrt{p(r)}$, they fall into the black hole singularity, see
Fig.~\ref{fig:case3b}.

\begin{figure}[htb]
  \centering
  \includegraphics[width=8cm]{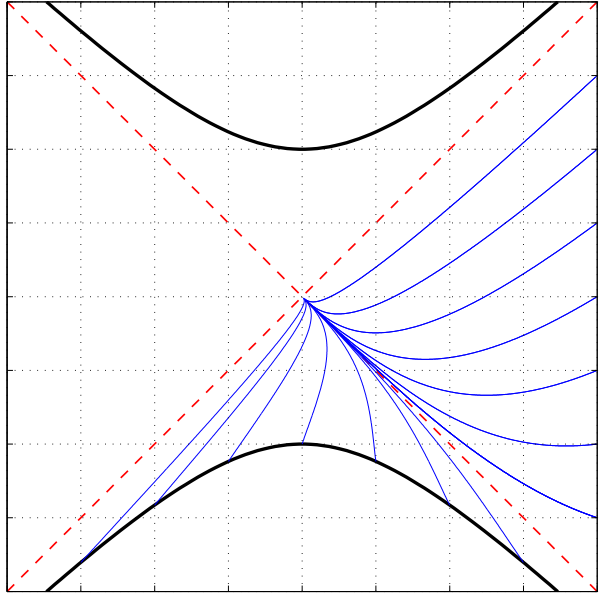}
  \caption{The case $c=1+a^2$ for $a^2=0.1$}
  \label{fig:case3c}
\end{figure}

Finally, in the critical case when $1+a^2=c \,$ the zeros of $p(r)$
and $q(r)$ coincide at $r=1$, the origin $T=0=X$ is a point where the
system~\eqref{eq:UVsystem} is singular.  This follows from the fact
that
\[
\sqrt{r p(r)} \pm \sqrt{r q(r)} = \sqrt{r-1} \left(
  \sqrt{r+a^2 (1+r+r^2)} \pm \sqrt{a^2 (1+r+r^2)}\right)
\, ,
\]
where the expression in parentheses is differentiable in $r$ near
$r=1$. Using the coordinate transformation between Schwarzschild and
Kruskal coordinates we see that in the region~I the system can be
written in the form
\begin{equation}
  \label{eq:UVsystemhoriz}
  \begin{aligned}
    \dot U &= \left\{ \sqrt{e} \left(\sqrt{3a^2}-\sqrt{1+3a^2}\right) +
    \text{reg}(U,V)\right\} \frac{\sqrt{-U}}{\sqrt{V}} \, , \\
    \dot V &= \left\{\sqrt{e} \left(\sqrt{3a^2}+\sqrt{1+3a^2}\right) +
    \text{reg}(U,V)\right\} \frac{\sqrt{V}}{\sqrt{-U}} 
\, ,
\end{aligned}
\end{equation}
where $\text{reg}(U,V)$ stands for a function which is regular near
$U=V=0$ and vanishes there. Near the origin the behaviour of the
solutions is given by this first exhibited term. Dropping the regular
term we arrive at the system
\begin{equation}
  \label{eq:UVsystemhorizreduced}
  \begin{aligned}
    \dot U &= \left\{ \sqrt{e}
      \left(\sqrt{3a^2}-\sqrt{1+3a^2}\right)\right\}
    \frac{\sqrt{-U}}{\sqrt{V}} \, , \\ 
    \dot V &= \left\{\sqrt{e}
      \left(\sqrt{3a^2}+\sqrt{1+3a^2}\right)\right\}
    \frac{\sqrt{V}}{\sqrt{-U}}
\, ,
\end{aligned}
\end{equation}
which has the solution
\begin{equation}
  \label{eq:solution}
  \begin{aligned}
    U(s) = U_0 \left(1 +
    \frac{\sqrt{1+3a^2}}{\sqrt{-U_0V_0}}s\right)^{1-\gamma} \, , \\
    V(s) = V_0 \left(1 +
    \frac{\sqrt{1+3a^2}}{\sqrt{-U_0V_0}}s\right)^{1+\gamma} \, ,
  \end{aligned} \qquad \gamma=\frac{\sqrt{3a^2}}{\sqrt{1+3a^2}}<1
\, .
\end{equation}
Following the hypersurface backward for $s<0$ from any initial point
$(U_0,V_0)$ we find that at $s=-\sqrt{-U_0V_0}/\sqrt{1+3a^2}$ both $U$
and $V$ become zero, the hypersurface runs into the origin and stops there. The
singularity at the origin therefore has the consequence that
uniqueness of the solution is lost, see~Fig.~\ref{fig:case3c}

\section{Discussion}
\label{sec:disc}

In this paper we have discussed the class of spherically symmetric
hypersurfaces with constant scalar curvature embedded in the
Schwarzschild space-time. We found several types of hypersurfaces with
qualitatively different behaviour. We have looked at this class of
hypersurfaces because we were interested in finding ways of slicing
the Schwarzschild space-time with hypersurfaces which become
asymptotically hyperboloidal, i.e. for which the scalar curvature
approaches a negative constant for large radii.

\begin{figure}[htb]
  \centering
  \includegraphics[width=8cm]{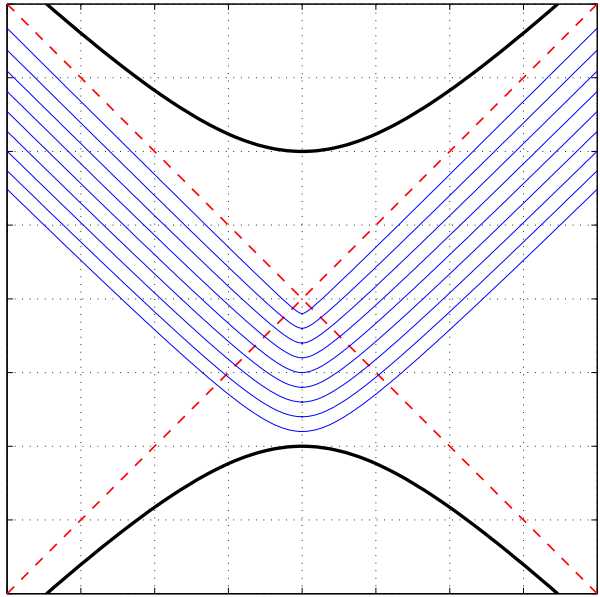}
  \caption{A set of symmetric hypersurfaces for fixed $a=1$ and
    equidistant intersection with the $T$-axis. 
  \label{fig:foliation}}
\end{figure}

We can now see that for the subclass of space-like hypersurfaces with $0<c<1+a^2$
one might be able to achieve this. All of these hypersurfaces 
connect the two future null-infinities of the Kruskal
extension. Hypersurfaces with $c<0$ could foliate the regions~I
and~III, i.e. the white hole and the asymptotic region.  However, note
that the hypersurfaces shown in Fig.~\ref{fig:case2} and
Fig.~\ref{fig:case3a} do not form a foliation because they intersect.
So one needs to vary the available parameters $a$ and $c$, as well as
the location of the hypersurface in the space-time, in order to avoid
intersections. To illustrate this we display in Fig.~\ref{fig:foliation}
several hypersurfaces with the fixed value of $a=1$ that are symmetric
under $X\mapsto -X$. So they have their minimal surface on the
$T$-axis. From the intersection of the hypersurfaces with this axis we
can determine the corresponding value of~$c$.  In this part of the
space-time the  family of hypersurfaces seems to give a foliation,
although we have not proved this. Discussions of these issues will be
left to another paper.

\section{Acknowledgments}
  
We are grateful to E.~Malec and N.~O'Murchadha for discussions and
hints. This work was supported in part by the Deutsche
Forschungsgemeinschaft (DFG).


\end{document}